\documentclass[10pt,epsf]{iopart}
\usepackage{epsf}
\usepackage{graphicx}
\usepackage{subfigure}
\newcommand{\be}{\begin{equation}}
\newcommand{\ee}{\end{equation}} 
\newcommand{\eei}{\end{equation}\indent\indent}
\newcommand{\bc}{\begin{center}}
\newcommand{\ec}{\end{center}}
\newcommand{\ber}{\begin{eqnarray}}
\newcommand{\ear}{\end{eqnarray}}
\newcommand{\ba}{\begin{array}}
\newcommand{\ea}{\end{array}}

\newcommand{\ti}{\tilde}

\newcommand{\bea}{\begin{eqnarray}}
\newcommand{\eea}{\end{eqnarray}}

\newcommand{\ei}{\end{itemize}}

\newcommand{\udot}{\dot{u}}

\newcommand{\bra}[1]{\left(#1\right)}

\newcommand{\nab}{\nabla}
\newcommand{\la}{\langle}
\newcommand{\ra}{\rangle}

\newcommand{\lb}{\{}
\newcommand{\rb}{\}}
\newcommand{\A}{{\cal A}}
\newcommand{\E}{{\cal E}}
\renewcommand{\H}{{\cal H}}
\newcommand{\LB}{\left(}
\newcommand{\RB}{\right)}
\newcommand{\LBB}{\left[}
\newcommand{\RBB}{\right]}
\newcommand{\lc}{\varepsilon}

\def\case#1/#2{\textstyle\frac{#1}{#2} }
\usepackage{graphicx}
\usepackage{epsfig}
\usepackage{rotating}
\begin{document}
\title{Gravitational Lensing in Spherically Symmetric Spacetimes}
\author{Bonita de Swardt$^2$, Peter K. S. Dunsby$^{1,2,3}$  and  Chris Clarkson$^{1,2}$}
\address{$1$ Centre for Astrophysics, Cosmology and Gravitation,  University of Cape Town, 7701 Rondebosch, South Africa.}
\address{$2.$ Department of Mathematics and Applied Mathematics,
  University of Cape Town, 7701 Rondebosch, South Africa.}
\address{$3.$ South African Astronomical Observatory, Observatory
7925, Cape Town, South Africa.}
\date{\today}
\begin{abstract}
We present a framework for studying gravitational lensing in spherically 
symmetric spacetimes using 1+1+2 covariant methods. A general formula 
for the deflection angle is derived and we show how this can be used to recover 
the standard result for the Schwarzschild spacetime.
\end{abstract}
\section{Introduction} \label{introduction}
In this paper, we apply the recently developed  1+1+2 covariant approach to the  lensing of  light rays in any spherically symmetric spacetime. A more detailed discussion of this approach can be found in \cite{bi:clarkson}, from which most of the material in section \ref{Background} has been drawn. We begin with a motivation for using the 1+1+2 approach rather than the conventional  1+3 covariant method used in a related paper by the same authors to derive the  cosmological lens equation \cite{bi:lensing1}. 

The 1+3 covariant approach \cite{bi:Covariant,bi:Perturbations} has proven to be a very powerful framework  in cosmology, not only in studying the motion of null rays and  gravitational lensing, but in many other areas of cosmology. In particular this approach  has been particularly useful in obtaining a deep understanding of many aspects of relativistic fluid flows,  particularly in areas involving perturbation methods. In cosmology these methods have been applied to the evolution of linear and non-linear fluctuations in the universe and to the physics of the cosmic microwave background \cite{bi:CMB}, amongst other things. The strength of this approach lies in the fact that all the essential information describing a particular system is captured in a set of 1+3 covariant variables which are defined relative to a preferred time-like congruence, often taken to be the 4-velocity $u^a$ of matter, allowing these variables to have an immediate physical and geometrical significance. These variables then satisfy a set of \textit{evolution} and \textit{constraint} equations which are derived from the Einstein Field Equations, the Bianchi and Ricci identities, forming a closed system of equations for a chosen equation of state describing the matter. 

A natural extension of the 1+3 approach involves a further splitting of the 3-space. This 1+1+2 decomposition of spacetime is ideal for solving problems which have spherical symmetry. In particular this approach was used to study linear perturbations of a Schwarzschild spacetime and to investigate the generation of electromagnetic radiation by gravitational waves interacting with a strong magnetic field in the vicinity of a vibrating Schwarzschild black hole \cite{bi:clarkson}.

In addition to the splitting with respect to a timelike vector $u^a$, this 1+1+2 approach relies on the further decomposition of  spacetime using a spacelike vector $e^a$. The Bianchi and Ricci identities can then be split using both $u^a$ and $e^a$ giving
rise to a set of coupled first order differential equations and constraints. Differential operators are along these two vector fields resulting in a set of evolution and propagation (along $e^a$) equations (see \cite{bi:clarkson} for the full set of equations). We also denote a differential operator orthogonal to the spacelike vector $e^a$ (and thus lying in the 2-dimensional sheet)  by~$\delta_a$.

The  1+1+2 decomposition of the null tangent vector $k^a$ gives more insight on the lensing geometry for a light ray traveling in a spherically symmetric spacetime and allows us to directly obtain the general form of the deflection angle $\alpha$, for spherically symmetric lensing situations, which is the main result of this paper.

Conventions are taken to be the same as in \cite{bi:lensing1} and unless otherwise stated, we  use geometrized units, so that $8\pi G=c=1$.

\subsection{The $1+1+2$ Covariant Sheet Approach} \label{Background}
In the 1+3 covariant approach, a 4-velocity field, $u^a= dx^a/d\tau$, is introduced which represents the average 4-velocity of a set of observers in the spacetime. The projection tensor, $h_{ab} = g_{ab}+u_{a}u_{b}$, can be used to project any vector or tensor orthogonal  to $u^a$. Thus, $h_{ab}$ allows for the decomposition of any 4-vector into a component parallel to $u^a$ and a component lying in the 3-space orthogonal to $u^a$.

Another vector field can be defined which allows us to perform an additional decomposition of the 1+3 equations. Introducing a unit spatial vector $e^a$ which lies orthogonal to the $u^a$ so that $u^ae_a = 0$ and $e^ae_a = 1$. The projection tensor 
\be 
N_a{}^b \equiv h_a{}^b - e_ae^b = g_a{}^b + u_au^b - e_ae^b~,~~N^a{}_a = 2~, \label{projT} 
\ee 
then projects vectors orthogonal to $e^a$ \textit{and} the timelike vector $u^a$, i.e. $e^aN_{ab} = 0 =u^aN_{ab}$. Thus, $N_{ab}$ projects a tensorial object onto 2-surfaces which we call the `sheet'. 

In the 1+3 covariant approach to the study of null congruences \cite{bi:lensing1}, we introduced the spatial vector, $n^a$, which was chosen such that $n^an_a = 1$ and $u^an_a = 0$ (i.e. $n^a$ is a unit spatial vector). The tensor, $\ti{h}^a{}_b \equiv h^a{}_b -n^an_b$, could then be defined as a projection tensor, which projects a vector or tensor into the 2-dimensional screen-space,
orthogonal to the null tangent vector $k^a$. In the following analysis, we have not allowed for 
the spatial vectors $n^a$ and $e^a$ to coincide. However, if we choose $n^a$ to coincide with $e^a$,
it immediately follows that $\ti{h}^a{}_b= N^a{}_b$, so that the screen and sheet represent the same 2-dimensional surface. 

Any 3-vector, $\psi^a$, can be irreducibly split into a component along $e^a$ and a sheet 
component $\Psi^a$, orthogonal to $e^a$ i.e. 
\be
\psi^a = \Psi e^a + \Psi^a~,~\Psi\equiv\psi^ae_a~{\mathrm
and}~\Psi^a \equiv N^{ab}\psi_b~. \label{equation1} 
\ee 
A similar decomposition can be done for a projected, symmetric, trace-free (PSTF) tensor, $\psi_{ab}$, which can be split into scalar, vector and tensor part as follows:
\be 
\psi_{ab} = \psi_{\la ab\ra} = \Psi(e_ae_b - \frac{1}{2}N_{ab})+2\Psi_{(a}e_{b)} + \Psi_{ab}~, 
\label{equation2} 
\ee 
where 
\bea
\Psi &\equiv & e^ae^b\psi_{ab} = -N^{ab}\psi_{ab}~,\nonumber \\ 
\Psi_a &\equiv & N_a{}^be^c\psi_{bc}~,\nonumber \\ \Psi_{ab}
&\equiv & \psi_{\left\{ab\right\}} \equiv \LB N^c{}_{(a}N_{b)}{}^d
- \frac{1}{2}N_{ab} N^{cd}\RB \psi_{cd}~, 
\eea 
and the curly brackets denote the PSTF part of a tensor with respect to $e^a$. We also have 
\be 
h_{\left\{ab\right\}} = 0~,~ N_{\la ab\ra} =
-e_{\la a}e_{b\ra} = N_{ab}- \frac{2}{3}h_{ab}~. 
\ee 
We now define the alternating Levi-Civita 2-tensor as 
\be
\lc_{ab}\equiv\lc_{abc}e^c = \eta_{dabc}e^cu^d~, \label{eq:perm}
\ee 
where $\lc_{abc}$ is the 3-space permutation symbol and
$\eta_{abcd}$ is just the spacetime permutator. With the
definition of $\lc_{ab}$ given by equation (\ref{eq:perm}) above,
we also have the following relations$~$\footnote{Note that for any
2-vector $\Psi^a$, we may use $\lc_{ab}$ to form a vector that
will lie orthogonal to $\Psi^a$ but of the same length.} : 
\bea
\lc_{ab}e^b &=& 0 = \lc_{(ab)}~, \\ \lc_{abc} &=& e_a\lc_{bc}
-e_{b}\lc_{ac} + e_c\lc_{ab}~, \\
\lc_{ab}\lc^{cd} &=&  N_a{}^cN_b{}^d - N_a{}^dN_b{}^c~,
 \\ \lc_a{}^c\lc_{bc} &=& N_{ab}~,~~ \lc^{ab}\lc_{ab} = 2~. 
 \eea 
From these definitions it follows that any object in the 1+1+2
setting can be split into scalars, 2-vectors in the sheet, and
PSTF 2-tensors (also defined in the sheet). These are the three
objects that remain after the splitting has occurred. In the 1+1+2
formalism we can introduce two new derivatives: 
\bea
\hat{\psi}_{a..b}{}
^{c..d} &\equiv & e^f\nab_f\psi_{a..b}{}^{c..d}~, \nonumber \\
\delta_f\psi_{a..b}{}^{c..d} &\equiv & N_a{}^f...N_b{}^gN_i{}^c
...N_j{}^dN_f{}^j\nab_j\psi_{f..g}{}^{i..j}~. 
\eea 
The hat-derivative represents the derivative along the $e^a$
vector-field. The $\delta$ -derivative is the projected derivative
onto the sheet\footnote{We note that one needs to project on every
free index when calculating the $\delta$-derivative.}.    

We are now able to decompose the covariant derivative of $e^a$ orthogonal to $u^a$ giving 
\be {\rm D}_ae_b = e_aa_b +
\frac{1}{2}\phi N_{ab} + \xi\epsilon_{ab} + \zeta_{ab}~, 
\ee 
where
\bea 
a_a &\equiv & e^c{\rm D}_ce_a = \hat{e}_a~, \\ \phi &\equiv &
\delta_ae^a~, \\  \xi &\equiv & \frac{1}{2}
\epsilon^{ab}\delta_ae_b~, \\ \zeta_{ab} &\equiv & \delta_{\lb a
}e_{b \rb }~.
\eea 
We see that when travelling along $e^a$,  $\phi$
represents the expansion of the sheet,  $\zeta_{ab}$ is the shear
of $e^a$ (i.e. the distortion of the sheet) and $a^a$ its
acceleration. We can also interpret $\xi$ as the vorticity
associated with $e^a$ so that it is a representation of the
`twisting' or rotation of the sheet. The full covariant derivative
of $e^a$ is 
\bea \nab_a e_b &=& - \A u_au_b - u_a\alpha_b + \LB
\Sigma + \frac13\Theta \RB e_a u_b + \LBB \Sigma_a  -
\lc_{ac}\Omega^c \RBB u_b \nonumber \\ &+& e_aa_b + \frac12\phi
N_{ab} + \xi\lc_{ab} + \zeta_{ab}~, \label{eq:FullDerive} 
\eea 
and
\bea 
\nab_au_b &=& -u_a\LB \A e_b + \A_b\RB + e_ae_b\LB
\frac13\Theta + \Sigma \RB + e_a\LB \Sigma_b + \lc_{bc}\Omega^c\RB
\nonumber \\&+& \LB \Sigma_a-\lc_{ac}\Omega^c \RB e_b + N_{ab}\LB
\frac13\Theta - \frac12\Sigma\RB + \Omega\lc_{ab}+\Sigma_{ab}~,
\label{eq:fullDerive} 
\eea 
for the decomposition of the 1+3 covariant derivative of $u^a$. These
derivatives will be used in later calculations. The derivative of
$e^a$ in the direction of $u^a$ is given by 
\be 
\dot{e}_a = \A u_a
+ \alpha_a~,~~{\mathrm{where} }~~\A = e^a\dot{u}_a~,
\ee 
and
$\alpha_a$ is the component lying in the sheet. The new variables
$a_a$, $\phi$, $\xi$, $\zeta_{ab}$ and $\alpha_a$ are fundamental
objects of the spacetime and their dynamics give us information
about the spacetime geometry. Essentially, they are treated on
the same footing as the kinematical variables of $u^a$ in the 1+3 approach. 

The covariant derivative of a scalar $\Psi$ is 
\be 
{\rm D}_a\Psi= \hat{\Psi}e_a + \delta_a\Psi~. 
\ee 
While for any vector $\Psi^a$ orthogonal to both $u^a$ and 
$e^a$ (i.e. $\Psi^a$ lies in the sheet), the various parts of its 
spatial derivative may be decomposed as follows 
\footnote{Note that a bar on a particular
index indicates that the vector or tensor lies in the sheet.} :
\be 
\fl
{\rm D}_a\Psi_b = -e_ae_b\Psi_ca^c + e_a\hat{\Psi} _{\bar{b}}
- e_b\LBB \frac{1}{2}\phi\Psi_a + \LB \xi\lc_{ac}+ \zeta_{ac}\RB
\Psi^c \RBB + \delta_a\Psi_b~. 
\ee 
Similarly, for a tensor
$\Psi_{ab}$ (where $\Psi_{ab} = \Psi_{\lb ab \rb}$) : 
\be 
\fl {\rm
D}_a\Psi_{bc} = -2e_ae_{(b}\Psi_{c)d}a^d + e_a\hat{\Psi}_{bc}
-2e_{(b}\LBB \frac{1}{2}\phi\Psi_{c)a} + \Psi_{c)}{}^d\LB
\xi\lc_{ad} + \zeta_{ad}\RB \RBB + \delta_a\Psi_{bc}~. 
\ee 
We include the following relations which may be useful: 
\bea
\dot{N}_{ab} &=& 2u_{(a}\dot{u}_{b)} - 2e_{(a}\dot{e}_{b)} =
2u_{(a}\A_{b)} -
2e_{(a}\alpha_{b)}~, \nonumber \\
\hat{N}_{ab} &=& -2e_{(a}a_{b)}~, \nonumber \\ \delta_cN_{ab} &=&
0~, \eea while 
\bea \dot{\lc}_{ab} &=& -2u_{[a}
\lc_{b]c}\A^c + 2e_{[a}\epsilon_{b]c}\alpha^c~, \nonumber \\
\hat{\lc}_{ab} &=& 2e_{[a}\lc_{b]c}a^c~,\nonumber \\
\delta_c\lc_{ab} &=& 0~. 
\eea 
We now split the (1+3) kinematical variables and Weyl tensors into the 
irreducible 1+1+2 set $\lb\Theta, \A, \Omega,\Sigma, \E, \H, \A^a, \Sigma^a, \E^a, \H^a,
\Sigma_{ab}, \E_{ab}, \H_{ab} \rb$ using equations
(\ref{equation1}) and (\ref{equation2}) giving: \bea \dot{u}^a &=&
\A e^a + \A^a~,\label{eq:1+1+2Acc} \\ \omega^a &=& \Omega e^a +
\Omega^a~,
\\ \sigma_{ab} &=& \Sigma\LB e_ae_b - \frac{1}{2}N_{ab}\RB +
2\Sigma_{(a}e_{b)} + \Sigma_{ab}~, \\ E_{ab} &=& \E\LB e_ae_b
- \frac{1}{2}N_{ab}\RB + 2\E_{(a}e_{b)} + \E_{ab}~, \\
H_{ab} &=& \H\LB e_ae_b - \frac{1}{2}N_{ab}\RB + 2\H_{(a}e_{b)} +
\H_{ab}~.\label{eq:MagH} 
 \eea 
We have now obtained the 1+1+2 form of the kinematical and Weyl 
variables corresponding to the irreducible parts of $\nab_ae_b$ (\ref{eq:FullDerive}). 
Similarly, we may split the anisotropic fluid variables $q^a$ and $\pi_{ab}$:
\bea
q^a &=& Qe^a + Q^a~, \\
\pi_{ab} &=& \Pi\LB e_ae_b -\frac12 N_{ab}\RB + 2\Pi_{(a}e_{b)} +
\Pi_{ab}~. \label{eq:Anisotropic}
 \eea

\section{Gravitational Lensing in Spherically Symmetric Spacetimes}
This section begins with a presentation of the various kinematical relations valid in the general local rotationally symmetric (LRS) spacetime. A detailed discussion of the covariant approach to LRS perfect fluid
spacetimes can be seen in \cite{bi:HvE}. However, in deriving the
lensing geometry we have restricted ourselves to the case of spherically 
symmetric spacetimes. We also derive the propagation equations for the 
lensing variables along $k^a$ in the general LRS spacetime. These equations 
will be needed in section \ref{Schw_defangle} where we derive the deflection angle in the
Schwarzschild spacetime.

\subsection{Kinematical Quantities}
As discussed in section \ref{introduction},  the 1+1+2 approach requires a
further splitting of spacetime variables, in addition to the
split with respect to the time-like vector $u^a$. This extra
splitting was done relative to the vector $e^a$. In the case of
LRS spacetimes, $e^a$ is just a vector pointing along the axis of
symmetry and can therefore be thought of as  a `radial' vector. LRS 
spacetimes are defined to be isotropic, allowing for the vanishing of \textit{all}
1+1+2 vectors and tensors, so that there are no preferred
directions in the sheet. The full covariant derivative of the
radial vector $e^a$ (\ref{eq:FullDerive}) and 4-velocity $u^a$
(\ref{eq:fullDerive}) reduce to  
\be 
\nab_a e_b = - \A
u_au_b + \LB \Sigma + \frac13\Theta \RB e_a u_b + \frac12\phi
N_{ab} + \xi\lc_{ab}~, \label{eRedCovariant} \ee and \be \nab_au_b
= -\A u_ae_b + e_ae_b\LB \frac13\Theta + \Sigma \RB + N_{ab}\LB
\frac13\Theta - \frac12\Sigma\RB + \Omega\lc_{ab}~,
\label{uRedCovariant} 
\ee 
respectively. We also find
\be 
\dot{e}_a = \A u_a~,\ee 
is the evolution of $e^a$. 

The kinematical quantities and Weyl tensors (\ref{eq:1+1+2Acc})
-(\ref{eq:Anisotropic})
reduce to  
\bea 
\dot{u}^a &=& \A e^a~, \\
\omega^a &=& \Omega e^a~,
\\ \sigma_{ab} &=& \Sigma(e_ae_b - \frac{1}{2}N_{ab})~, \\
q^a &=& Qe^a~, \\ \pi_{ab} &=& \Pi\left(e_ae_b - \frac12 N_{ab} \right)~, \\
E_{ab} &=& {\cal E}( e_ae_b - \frac{1}{2}N_{ab})~, \\ H_{ab} &=&
{\cal H}(e_ae_b - \frac{1}{2}N_{ab})~,  
\eea 
in an LRS spacetime where we have not made the perfect 
fluid assumption (i.e. $Q \neq \Pi \neq 0$ ).

We now list the set of 1+1+2 equations resulting in any
LRS (and thus any spherically symmetric) spacetime: 
\bea 
\dot\phi &=&
\bra{\frac23\Theta-\Sigma}\bra{\A-\frac12\phi} +2\xi\Omega+Q~,
\label{eq1}
\\
\dot\xi &=& \bra{\frac12\Sigma-\frac13\Theta}\xi
+\bra{\A-\frac12\phi}\Omega+\frac12\H~,
\\
\dot\Omega&=&+\A\xi+\Omega\bra{\Sigma-\frac23\Theta}~,
\\
\hat \A&=&\frac13\dot\Theta - \dot\Sigma
-\A^2+\bra{\frac13\Theta+\Sigma}^2
        +\frac16\bra{\mu+3p-2\Lambda}  +\E 
        -\frac12\Pi~,
\\ 
\hat\A&=&\dot\Theta-\bra{\udot+\phi}\A+\frac13\Theta^2
    +\frac32\Sigma^2-2\Omega^2
    + \frac12\bra{\mu+3p}  -\Lambda~,
\\
\dot\Sigma&=&\frac23\hat\A+\frac13\bra{2\A-\phi}\udot-\bra{\frac23\Theta
+\frac12\Sigma}\Sigma
     - \frac23\Omega^2-\E +\frac12\Pi~,
\\
\dot\mu&=&-\hat Q-\Theta\bra{\mu+p}-\bra{\phi+2\A}Q -
    \frac32\Sigma\Pi~,
\\
\dot Q&=&-\hat
p-\hat\Pi-\bra{\frac32\phi+\A}\Pi-\bra{\frac43\Theta+\Sigma} Q
    -\bra{\mu+p}\A~,
\\
\dot\E&=&-\frac12\dot\Pi-\frac13\hat Q
    +\bra{\frac32\Sigma-\Theta}\E
    -\frac12\bra{\frac13\Theta+\frac12\Sigma}\Pi\nonumber\\&&
    +\frac13\bra{\frac12\phi-2\A}Q
    +3\xi\H
    -\frac12\bra{\mu+p}\Sigma~,
\\
\dot\H&=&-3\xi\E
    +\bra{\Theta+\frac32\Sigma}\H+\Omega Q+\frac32\xi\Pi~,
\\
\hat\phi&=&-\frac12\phi^2+2\xi^2+\bra{\frac13\Theta
+\Sigma}\bra{\frac23\Theta-\Sigma}\nonumber\\
    &-&\frac23\bra{\mu+\Lambda} -\frac12\Pi-\E~,
\\
\hat\xi&=&-\phi\xi+\bra{\frac13\Theta+\Sigma}\Omega~,
\\
\hat\Sigma&=&\frac23\hat\Theta-\frac32\phi\Sigma-2\xi\Omega-Q~,
\\
\hat\Omega&=&+\bra{\A-\phi}\Omega~,\label{hatOmSnl}
\\
\hat\E&=&\frac13\hat\mu-\frac12\hat\Pi+
    -\frac32\phi\bra{\E+\frac12\Pi} +\frac12\Sigma Q+3\Omega\H~,
\\
\hat\H&=&
    -\frac32\phi\H-\bra{3\E+\mu+p-\frac12\Pi}\Omega-Q\xi~.
\\
0&=&\bra{2\A-\phi}\Omega
    -3\xi\Sigma+\H \label{eq2}
\eea

\subsection{Lensing Geometry}
In \cite{bi:lensing1} we introduced the spatial vector, $n^a$, which is a
3-vector in the direction of light propagation. Allowing for a
further split of this vector with respect to $e^a$ gives : \be n^a
= \kappa e^a + \kappa^a~.\label{3-vector}\ee Thus, $\kappa$ is the
magnitude of the radial component and $\kappa^a$ is the component
lying in the 2-dimensional sheet. In this way we are able to write the
null tangent vector $k^a$ as \be k^a = E(u^a
+ \kappa e^a + \kappa^a)~,\label{ReducedNull} \ee
where $E=-k^au_a$ is the energy of a photon relative to $u^a$.
\begin{figure}[htbp]
\begin{center}
\includegraphics[scale=0.7]{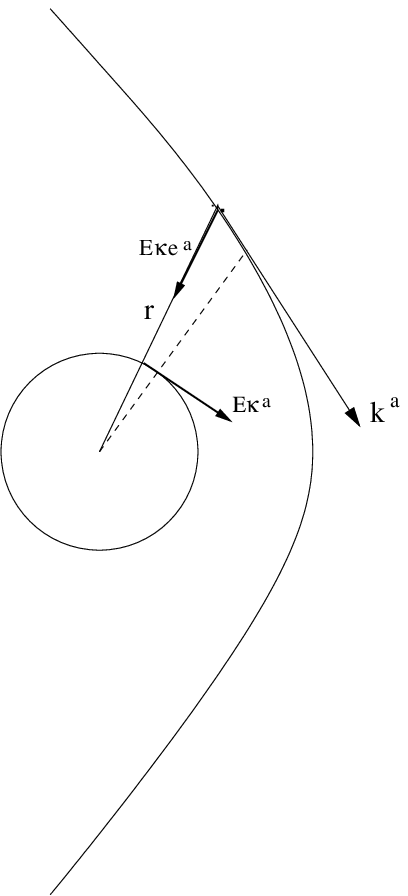}
\includegraphics[scale=0.7]{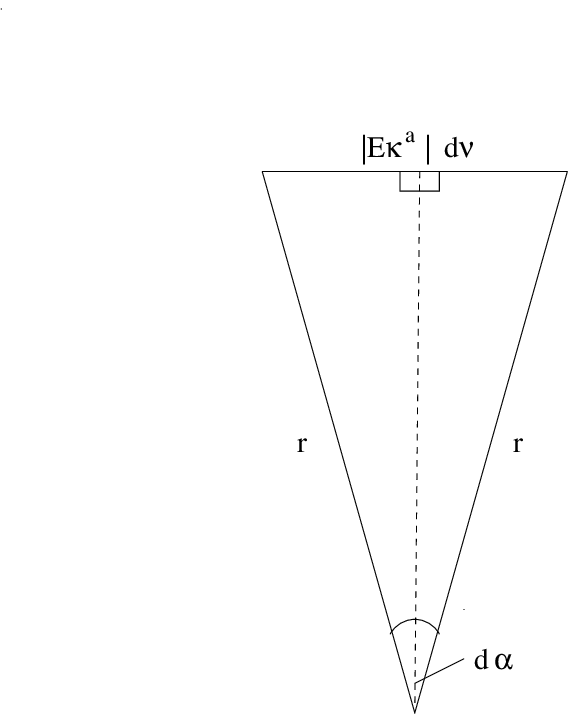}
\end{center}
\caption{\textbf{Lensing geometry in symmetric spacetimes}}
\end{figure}
Using this form of the null vector, we can determine the lensing geometry of a photon experiencing a deflection about the center of symmetry.
\subsection{General Form of the Deflection Angle}
Figure 1a traces the path of a photon in the presence of a strong
gravitational field such as a black hole, with $k^a$ lying tangent
to the null geodesic. We can also see the directions of the
various vector components of the null vector. The 2-dimensional
sheet is represented as a sphere about a central point. 
When considering infinitesimal small angles of deflection
$d\alpha$, the lensing geometry will be given by figure 1b. The
angle $d\alpha$ subtends an infinitesimal small displacement of
the photon's path, given by $\left|E\kappa^a\right|$. $d\alpha $
can be obtained from figure~1b giving  
\be 
d\alpha =
\frac{1}{r}\left|E\kappa^a\right|d\nu~,\label{DEFangle}\ee which
may be integrated giving the total deflection angle as \be \alpha
= \int_{\nu_{1}}^{\nu_{2}}{\frac{1}{r}\left|E\kappa^a\right|}d\nu
- \alpha_0~, \label{eq:ScalarAngle} \ee where $\alpha_0$ is the
integration constant. This is a completely general form of the
\textit{scalar deflection angle} which can be applied to any
spherically symmetric spacetime. The deflection angle $\alpha$
(\ref{eq:ScalarAngle}) may be written as \be \alpha =
\int_{\nu_{1}}^{\nu_{2}}{\frac{1}{r}\left|E\right|
\sqrt{\kappa^a\kappa_a}} d\nu - \alpha_0~, 
\ee 
where the magnitude
of $\kappa^a$ (i.e. $\left|\kappa^a\right|$) is just
$\sqrt{\kappa^a\kappa_a}$. Substituting for $k^a$
(\ref{ReducedNull}) into the null condition, $k_ak^a = 0$, gives a
restriction on $\kappa^a$ :
\be 
\kappa_a\kappa^a
=1-\kappa^2~,\label{KappaEqn}\ee 
so that the general deflection angle takes the form: 
\be 
\alpha =
\int_{\nu_{1}}^{\nu_{2}}{\frac{1}{r}\left|E\right| \sqrt{1 -
\kappa^2}}d\nu - \alpha_0~. \label{SchwScalarAngle} 
\ee 
Thus, if we know the relations $E(\nu)$, $\kappa(\nu)$ and $r(\nu)$, for a
given spherically symmetric spacetime, it is possible to find an
explicit form of the deflection angle. The differential equations
for these variables in the direction of the ray are derived in the next section.
\subsection{General Propagation Equations along $k^a$}
We begin this section by deriving the propagation equations for the
lensing variables $E$ and $\kappa$ in the direction of the null
ray. Each ray is parameterized by the affine parameter, $\nu$, so
that the geodesic condition can be written as :
 \be k^b\nab_bk^a =
\frac{\delta k^a}{ \delta\nu} = (k^a)' = 0~,\label{NullCond1} 
\ee
where the prime derivative ($'$) denotes change along the ray
(i.e. with respect to the affine parameter $\nu$). The geodesic
condition (\ref{NullCond1}) can be used to determine the propagation
equations for $E$ and $\kappa$. Substituting for the null vector
$k^a$ (\ref{ReducedNull}) into (\ref{NullCond1}) gives 
\bea 
\fl
\LBB
E' + E^2\kappa\A + E^2\kappa^2\LB \Sigma + \frac13\Theta \RB \RBB
u_a + \LBB E'\kappa + E^2\A + E^2\kappa\LB \frac13\Theta + \Sigma
\RB  + E\kappa{}'\RBB e_a \nonumber \\ \fl+ \LBB E' +
\frac{1}{2}E^2\phi + E^2\LB \frac13\Theta - \frac12\Sigma \RB \RBB
\kappa_a + E^2\LB \Omega + \xi \RB\lc_{ab}\kappa^b + E\kappa_a' =
0~,  \label{eq:project1}
 \eea 
where we have also
used 
\bea
 \fl u_a' &=& E\A e_a + E\kappa\LB \frac13\Theta + \Sigma \RB
e_a + E\LB \frac13\Theta
- \frac12\Sigma \RB\kappa_a + E\Omega\lc_{ab}\kappa^b ~,\label{eq:uPrime} \\
\fl e_a' &=& E\A u_a + E\kappa\LB \Sigma + \frac13\Theta \RB u_a +
\frac12E\phi\kappa_a + E\xi\lc_{ab}\kappa^b~, \label{eq:ePrime}
\eea
which are obtained from (\ref{eq:fullDerive}) and
(\ref{eq:FullDerive}), respectively. When (\ref{eq:project1}) is
projected along the timelike direction ($u^a$) and radial
direction ($e^a$), we obtain the general equations of $E'$ and
$\kappa'$, respectively. These are : 
\bea E' &=& -E^2\kappa\A -
2E^2\kappa^2\Sigma - E^2\LB \frac13\Theta - \Sigma\RB~,
\label{eq:Eprime1}\\
\kappa' &=& -E\LB 1-\kappa^2\RB \LB \A - \frac{1}{2}\phi +
2\kappa\Sigma \RB~, \label{eq:Kprime1}
 \eea 
 which can be used in
a general LRS spacetime. \\
\\The LRS class II spacetime \cite{bi:HvE}, admits spherically symmetric
solutions. This spacetime is free of rotation allowing for the
vanishing of the variables $\H$, $\Omega$ and $\xi$. If in
addition we describe the matter to be a perfect fluid so that $Q =
\Pi = 0$, then we can determine the propagation of some of the
scalars, $\left\{\mu,~ p,~\Theta,~\Sigma,~\A,~\E,~\phi \right\}$,
along $k^a$ using\footnote{Note that $\delta_aX = 0$ in a
spherically symmetric spacetime.} 
\be 
X' = E \LB \dot{X} +
\kappa\hat{X} \RB~, 
\ee 
where $X$ can be found through the
addition and subtraction of the relevant 1+1+2 equations
(\ref{eq1})-(\ref{eq2}) \footnote{We also note that in using a
spherically symmetric spacetime, these equations will reduce to
simpler expressions as $\H = \Omega = \xi = 0$ and $Q = \Pi =0$.}.

\section{Application to Lensing in Schwarzschild Spacetimes}

Schwarzschild gravitational lensing in a weak gravitational field (i.e. where the deflection angle is small) is well known (see for example \cite{bi:Weinberg}). An extension of Schwarzschild lensing in the  strong field limit is investigated in 
\cite{bi:EllisV2000} and \cite{bi:Bozza2002}.   

We have general forms for the deflection angle and the propagation
equations of the lensing variables along $k^a$. These will now be
applied to the Schwarzschild spacetime in which we hope to
recover the standard lensing results found in \cite{bi:Weinberg}.
\subsection{Solutions for $\A$, ${\cal E}$ and $\phi$}

The Schwarzschild spacetime is characterized as being static,
spherically symmetric and a vacuum. This covariant characterization
of the spacetime allows for the quantities $\A$, ${\cal E}$ and
$\phi$ to be non-zero, with all the other covariant quantities
vanishing\footnote{This covariant characterization of the 1+1+2
variables in a Schwarzschild spacetime is also discussed in
\cite{bi:clarkson}.}. The remaining propagation
equations are \cite{bi:clarkson} 
\bea 
\hat{\phi} &=& -\frac12\phi^2 - {\cal E}~, \label{eq:phiHat} \\
\hat{\cal E} &=& -\frac32\phi{\cal E}~, \label{eq:eHat} 
\eea
together with an extra equation 
\be {\cal E} + \A\phi = 0~.
 \ee
These are a closed set of equations from which we are able to find
solutions for the variables $\A$, $\phi$ and ${\cal E}$. The
coupled differential equations seen in (\ref{eq:phiHat}) and
(\ref{eq:eHat}) can be solved explicitly when the hat-derivative
is associated with an affine parameter, say $\rho$, so that
$\hat{} = d/d\rho$. The parametric solutions are :
\bea 
{\cal E}&=&-\frac{1}{(2m)^2}{\mathrm sech^6} x=-\frac{2m}{r^3}~,\label{eq:solE}\\
\phi &=& \frac{1}{m}{\mathrm sinh}~x~{\mathrm
sech^3}~x=\frac{2}{r}\sqrt{1-\frac{2m}{r}}~,
\label{eq:solPhi}\\
\A &=& \frac{1}{4m}{\mathrm cosech}~x~{\mathrm sech^3}~x =
\frac{m}{r^2}\left(1-\frac{2m}{r}\right)^{-1/2}~, \label{eq:solA}
\eea 
where solutions can be written in terms of either parameter
$x$ or $r$, and are related to the affine parameter through the
relation \be \rho = 2m\left( x + {\mathrm sinh}~x~{\mathrm cosh}~x
\right)~. \ee $r$ is the usual Schwarzschild radial coordinate
related to parameter $x$ by \be r = 2m~{\mathrm cosh^2}~x~. \ee
Solutions (\ref{eq:solE}) - (\ref{eq:solA}) are a one-parameter
family of solutions, parameterized by the Schwarzschild mass $m$.
Note that the Schwarzschild solution is given for $2m < r <
\infty$ for $0 < x < \infty$.

\subsection{Solutions for the Lensing Variables}

As stated in the previous section, the only non-zero 1+1+2
variables in a Schwarzschild spacetime are the scalars $\left\{\A,
\E, \phi \right\}$ (and their derivatives $\{\hat{\A}, \hat{\E},
\hat{\phi} \}$). Thus, the general propagation equations of $E$
(\ref{eq:Eprime1}) and $\kappa$ (\ref{eq:Kprime1}), in the
direction of the ray, reduce to the form : 
\bea 
E{}' &=&
-E^2\A\kappa~,
\label{eq:Eprime}\\
\kappa{}' &=& E(1-\kappa^2)(\frac{1}{2}\phi-\A)~.
\label{eq:Kprime} 
\eea 
Equation (\ref{KappaEqn}) gives the
propagation equation for $\kappa^a$ when differentiated with
respect to $\nu$ : 
\be \kappa^a{}' = -E\kappa\left(\frac{1}{2}\phi
- \A\right)\kappa^a - \frac12E\phi\left( 1 - \kappa^2 \right)e^a~.
\label{eq:K^aprime} 
\ee 
In order to find solutions to these
differential equations, we also include the propagation equations
for $\A$, ${\cal E}$ and $\phi$. These are obtained by
substituting for the 1+1+2 form of $k^a$ (\ref{ReducedNull}) into
the definition of the prime derivative of a particular variable.
We then find:
\bea
\A{}' &=& k^a\nab_a\A = -E\kappa(\phi + \A)\A~,\label{eq:Aprime}\\
\phi{}' &=& k^a\nab_a\phi = E\kappa(\A-\frac{1}{2}\phi)\phi~,\label{eq:Phiprime}\\
{\cal E}{}' &=& k^a\nab_a{\cal E} = -\frac{3}{2}E\kappa\phi{\cal
E}~.\label{eq:E'} 
\eea 
The differential equations
(\ref{eq:Aprime})-(\ref{eq:E'}) can be used in solving for $E$ and
$\kappa$ as will be described below. 

Dividing equation (\ref{eq:Phiprime}) by $\phi$ gives:
 \be
\frac{\phi{}'}{\phi} = E\kappa\A - \frac12 E\kappa\phi~.
\label{eq:step1} 
\ee 
We can then use the differential equations
for $E$ (\ref{eq:Eprime}) and $\E$ (\ref{eq:E'}) to rewrite
(\ref{eq:step1}) as:
 \be 
 \frac{E{}'}{E} = -\frac{\phi{}'}{\phi} +
\frac13\frac{\E{}'}{\E}~,
 \ee which can be solved as a total
differential equation, i.e. 
\bea 
\LB {\mathrm ln}~E \RB' &=& \LB
{\mathrm ln}~\phi - {\mathrm ln}~\E^{1/3} \RB'
= \LB {\mathrm ln}~\frac{\phi}{\E^{1/3}} \RB' \nonumber \\
\Rightarrow E &=& \frac{\phi}{C_1{\E}^{1/3}}~,
\label{eq:SolEnergy} 
\eea 
with $C_1$ being an integration constant
with respect to affine parameter $\nu$. Similarly, we are able to
solve for $\kappa$ starting with (\ref{eq:Phiprime}) and using the
differential for $\kappa$ (\ref{eq:Kprime}). We find: 
\bea
\frac{\kappa\kappa{}'}{1-\kappa^2} = -\frac{\phi{}'}{\phi}~, 
\eea
which can be written in total differential form as 
\be 
\LBB
\frac12{\mathrm ln}~\LB1-\kappa^2\RB \RBB' = \LB {\mathrm ln}~\phi
\RB{}'~.\ee This equation can be integrated giving \be \kappa =
\LB 1-\frac{\phi^2}{C_2}\RB ^{\frac{1}{2}}~, \label{eq:SolK} 
\ee
where
$C_2$ is a constant of integration with respect to $\nu$. 

We can now rewrite the solutions for $E$ and $\kappa$ in
terms of the radial coordinate $r$ and Schwarzschild mass $m$. In
order to do this, the integration constants $C_1$ and $C_2$ need
to be determined. $C_1$ can be obtained by taking the limit of
equation (\ref{eq:SolEnergy}) as $r$ tends to $\infty$, and given
the solutions of ${\E}$ and $\phi$ from equations (\ref{eq:solE})
and (\ref{eq:solPhi}), respectively. We find
\be 
C_1 =
-\frac{(2m)^{1/3}}{2E_{\infty}}~, \label{eq:C1}
\ee 
where we denote the energy at a large radial distance as, $E_{\infty}$. 

At the point of closest approach \footnote{Here $r_0$ is just the closest  point that the ray would reach in the vicinity of the lensing object.}, $r =r_0$, we must have $\kappa=0$ so that 
we are able to determine $C_2$ from equation (\ref{eq:SolK}) above. We obtain : 
\be 
C_2 =
\sqrt{\frac{4(r_0-2m)}{r_0{}^3}}~, \label{eq:C2} 
\ee 
where $r_0$
is the {\it distance of closest approach}${}$. Substituting for
the constants of integration (\ref{eq:C1}) and (\ref{eq:C2}), into
(\ref{eq:SolEnergy}) and (\ref{eq:SolK}) respectively, gives :
\bea 
E &=& E_{\infty}\left(1
- \frac{2m}{r}\right)^{-\frac{1}{2}}~, \label{EnergySol2} \\
\kappa &=& \pm\left[1 -
\frac{r_0^3}{r^2(r_0-2m)}\left(1-\frac{2m}{r}\right)\right]^{
\frac{1}{2}}~,\label{kappaSol2} 
\eea 
Equations (\ref{EnergySol2})
and (\ref{kappaSol2}), together with the solutions for $\left\{
\A,~\E~\phi \right\}$, form a complete set of solutions from which
the lensing geometry in a Schwarzschild spacetime can be
determined. This is described in the next section.

\section{The Deflection Angle $\alpha$} \label{Schw_defangle}
We would like to rewrite the scalar deflection angle (\ref{SchwScalarAngle}) in terms of the Schwarzschild mass $m$ and
radial coordinate $r$ so that it may be compared to that given in the standard lensing literature.  Substituting for the solutions for $E$ (\ref{EnergySol2}) and $\kappa$ (\ref{kappaSol2}) into (\ref{SchwScalarAngle}) gives 
\bea 
\alpha
&=& \int_{\nu_{1}}^{\nu_{2}}{
\frac{E_{\infty}}{r}\left(\frac{r_0^3}{r^2(r_0-2m)}\right)^{1/2}}
d\nu - \alpha_0 \nonumber \\ &=& \int_{\nu_{1}}^{\nu_{2}}{
E_{\infty}\frac{J}{r^2}}d\nu - \alpha_0~,
\label{SchwScalarAngle1}
\eea 
where $J$ is the impact parameter defined as 
\be
 J = r_0\left(1 -
\frac{2m}{r_0}\right)^{-1/2}~.\label{eq:ImpactP} 
\ee 
We need a transformation relation between the affine parameter $d\nu$ and
radial distance $dr$. Differentiating the solution for $\E$
(\ref{eq:solE}) with respect to $\nu$ gives: 
\be
 \E{}' =\frac{6m}{r^4}r{}'~,\label{eq:E'1} 
 \ee 
 and substituting for
$\kappa$ (\ref{kappaSol2}), $\phi$ (\ref{eq:solPhi}) and $\E$
(\ref{eq:solE}) into the differential equation for $\E$
(\ref{eq:E'}) results in 
\be
 \E{}' = E\frac{6m}{r^4}\LB 1 -
\frac{2m}{r} \RB^{1/2} \LBB 1 - \frac{r_0{}^3}{r^2(r_0 - 2m)}\LB 1
- \frac{2m}{r} \RB \RBB^{1/2}~.\label{eq:E'2} 
\ee 
The transformation relation can be obtained in equating equations
(\ref{eq:E'1}) and (\ref{eq:E'2}) giving
 \bea 
 dr &=&
E_{\infty}\left[ 1 - \frac{r_0{}^3}{r^2\left( r_0 - 2m
\right)}\left(1 - \frac{2m}{r}\right) \right]^{1/2}d\nu~,
\nonumber
\\ &=& E_{\infty} \left[ 1 - \frac{J^2}{r^2}\left(1 -
\frac{2m}{r}\right) \right]^{1/2}d\nu~. 
\eea 
Using this transformation in equation (\ref{SchwScalarAngle1}), gives the
form of the scalar deflection angle in terms of the impact
parameter $J$ and radial coordinate $r$ as : 
\be 
\alpha =
2\int_{r_0}^{\infty}{ \frac{J}{r^2}\left[ 1 -
\frac{J^2}{r^2}\left(1 - \frac{2m}{r}\right) \right]^{-1/2} }dr -
\alpha_0~,\label{eq:AAngle} 
\ee 
which is the exact expression
obtained in the lensing literature \cite{bi:Weinberg} where a
metric based approach is used (and if we take $\alpha_0 = \pi$).

\section{Conclusions}

In this paper we have developed a new approach for studying lensing effects in spherically symmetric 
spacetimes based on the 1+1+2 covariant approach to general relativistic fluid dynamics. In particular, 
we derived a completely general formula for the bending angle which is expressed in terms of the 
geometry of the null congruence. This approach proves particularly useful in the study of  lensing in 
$f(R)$ theories of gravity \cite{bi:Nzioki}.

\section*{References}

\end{document}